\documentclass[a4paper,16pt]{article}
\usepackage{graphicx}

\font\bfdodc=cmbx12 

\def\p{^{^{\prime}}}\def\pp{^{^{\prime\prime}}}
\def\ppp{^{^{\prime\prime\prime}}}
\def\ppp{^{^{\prime\prime\prime}}}

\def\bq{{\bf q}} 

\def\bq{{\bf q}}
\def\bfr{{\bf r}}\def\bR{{\bf R}}
\def\begeq{\begin{equation}}
\def\endeq{\end{equation}}
\def\begdis{\begin{displaymath}}
\def\enddis{\end{displaymath}}

\def\cA{{\cal A}}\def\cB{{\cal B}}\def\cC{{\cal C}}\def\cD{{\cal D}}
  
\def\cQ{{\cal Q}}  
  
\def\cS{{\cal S}}
                        
\def\cQS1{{\cQ_{\cS_1}}}

\def\ome#1#2{\hat{\Omega}(h_#1,r,#2)}
\def\odme#1#2{\hat{\Omega}_1(h_#1,r,#2)}

\def\eg{{\em e.g.}}

\def\hw{{\hat \omega}}

    \def\Aang{{\cal A}(\alpha,\beta,\gamma)}
     \def\Bang{{\cal B}(\alpha,\beta,\gamma)}
     \def\Cang{{\cal C}(\alpha,\beta,\gamma)}

\def\ac{\hfill\break\noindent }

\def\href#1{\relax}
\begin{document}

\title{The isotropic correlation function of plane figures: the triangle
case.}

\author{
{{Salvino Ciccariello}}
\\[6mm]
\begin{minipage}[t]{3mm}
           \
\end{minipage}
\begin{minipage}[t]{0.9\textwidth}
  \begin{flushleft}
  \setlength{\baselineskip}{12pt}
  {\slshape{\footnotesize{
  Universit\`{a} di Padova, Dipartimento di Fisica {\em G. Galilei}
  }}}\\{\slshape{\footnotesize{
  Via Marzolo 8, I-35131 Padova, Italy.
  }}}\\{\slshape{\footnotesize{
  E-mail {\upshape{\texttt{salvino.ciccariello@pd.infn.it}}}; Phone +39~049~8277173
  ;\\ Fax +39~049~8277102
  }}}
  \end{flushleft}
\end{minipage}\\[10mm]
}

\date{June 5, 2010}

\maketitle
\begin{abstract}
The knowledge of the isotropic correlation function of a plane
figure is useful to determine the correlation function of the
cylinders having the plane figure as  right-section and a given height
as well as  to analyze the out of plane intensity collected in grazing
incidence small-angle scattering from a film formed by a particulate
collection of these cylinders.
The correlation function of plane polygons can always be determined
in closed algebraic form. Here we report its analytic expression
for the case of a triangle. The expressions take four different forms
that depend on the relative order among the sides and the heights
of the triangle.

\bigskip
\centerline{May 26, 2010}
\noindent Math  Rev Class Numb: 78A45, 60D05, 52A22, 53C65\\

\hfill{DFPD 10/TH/10\quad}
\end{abstract}
\section{Introduction}
In small-angle scattering (SAS) theory samples are conceived of being made
up of two/three homogeneous phases\cite{guinier55,debye57}. In this way
the collected scattering intensities are used to determine few structural
parameters of the observed samples. In principle the procedure is very
simple. Restricting ourselves for definiteness to the case of a two-phase
particulate sample and assuming that the particle shape is unique, then the
effective scattering density of the sample reads
\begeq\label{1.1}
n(\bfr)=(n_1-n_2)\sum_{j=1}^N \rho\bigl({{\cal R}^{-1}}_{{\hat \Omega}_j}
(\bfr/D_j)-\bR_j\bigr),
\endeq
where $(n_1-n_2)$ is the contrast and the sum is performed over the $N$
particles of the samples. Moreover,  $D_j$, $\bR_j$ and ${{\hat \Omega}_j}$
respectively denote the size, the  mass-centre position of the $j$th particle
and the three Euler angles specifying  the rotation which the $j$th particle
has undergone to. ${\cal R}_{{\hat \Omega}_j}$ denotes the corresponding
rotation matrix.  Finally, $\rho(\bfr)$ is the function that specifies the
particle shape since it is equal to one or zero depending on whether the
tip of $\bfr$ falls inside or outside the particle of unit size ($D=1$)
with its centre (of mass) set at the origin. The observed scattering
intensity $I(\bq)$ $\bigl[$with $q=|\bq|=(4\pi/\lambda)\sin(\theta/2)$,
$\lambda$ denoting the ingoing beam particle wave-length and $\theta$ the
scattering angle$\bigr]$ is simply proportional to $|{\tilde n}(\bq)|^2$,
the square modulus of the Fourier transform (FT) of Eq.~(\ref{1.1}). The
fit of the observed $I(\bq)$ to $|{\tilde n}(\bq)|^2$, calculated  by
the procedure sketched below, allows
one to fix the values of the parameters present in
Eq.~(\ref{1.1}). In practice, before proceeding to the best fit analysis,
it is necessary to exploit any information on the physical
structure of the sample. To this aim one first considers the convolution
of $n(\bfr)$ and then uses the property that its FT yields the scattering
intensity. The convolution involves the double sum  $\sum_{j,l=1}^N T_{j}*
T_{l}$ where $T_{j}$ denotes the $j$th addend of Eq.~(\ref{1.1}).  In
handling the sum, it is convenient to separate  isotropic from anisotropic
samples. In the first case, the sum of the terms involving particles with
different orientations is assumed to average to zero provided we substitute
the remaining convolutions with their isotropic components.
These, in turn, are obtained by suitably scaling the expression
\begeq\label{1.2}
\gamma_3(r)\equiv \frac{1}{4\pi V}\int d\hw \int\rho(\bfr_1)
\rho(\bfr_1+r\hw)dv_1,
\endeq
which defines the isotropic (auto)correlation function (CF) of the
particle of unit size and volume $V$.  The FT of $\gamma_3(r)$ is the
particle (isotropic) form-factor. The scattering intensity is
\begeq\label{1.3}
I(q) \propto \sum_{j}D_j^6{\tilde\gamma}_3(D_j q)+\sum_{j\ne l}D_j^3
D_l^3{\tilde\gamma}_3(D_j q)
{\overline{{\tilde\gamma}_3(D_l q)}}e^{-i\bq\cdot(\bR_j-\bR_l)},
\endeq
where the tilde and the over-bar respectively denote the FT and the
complex conjugate. Further approximations as the decoupling
\cite{guinier55} and the local monodisperse one \cite{peders97} allow us
to simplify somewhat more expression (\ref{1.3}). In any case, the
final expression requires the knowledge of ${\tilde\gamma}_3(D_j q)$ which
is algebraically known only for the spherical shape. It can numerically be
evaluated by a one-dimensional integral for the parallelepiped\cite{good80,
gille87}, the circular cylinder\cite{cicc91}, the rotational
ellipsoid\cite{burgrul} and the tetrahedron\cite{cicc05} because the CFs
of these shapes are known in terms of transcendental functions. In all the
other cases it requires a 3-5 dimensional integral, a condition that makes
any numerical analysis with other particle shapes much more awkward.
Recently it has been shown that the CF of a cylinder of arbitrary right
section is related to the CF of the plane section by a generalized Abel
transform\cite{gille02,cicc03}. Thus, if one algebraically knows the
correlation function of a plane figure, the CF of the associated cylinder
is obtained by a quadrature and its FT by a further integration.
Generalizing the fact that the CFs of the regular triangle,
square\cite{sulank}, pentagon\cite{harut} and hexagon\cite{aharon}
are algebraically known\footnote{\label{ddd1}{Very recently it has been
determined the algebraic expression of the CF of a regular polygon whatever
the side number\cite{ohanyan}.}}, Ciccariello\cite{cicc09} has recently
shown that the CF of any plane polygon can be determined in closed
algebraic form. By this result it is now practically possible to consider
cylindrical shapes more involved than the circular ones,
see \eg\ \cite{gille09}.\ac
Surprisingly the algebraic expression of the CF of the simplest polygonal
shape, namely that of a generic triangle, as yet is not known.  In this
paper we fill up the gap. This will be done
in the next section.  Before concluding this section we wish to emphasize
that the mentioned property of the CF of a plane polygon can also be used in
the case of some anisotropic samples. In fact, in the case of fiber samples
(see, \eg, Ref.\cite{burger}), one uses the anisotropic CF of a circular
cylinder. By the aforesaid result it becomes now possible to consider the
correlation function of a polygonal cylinder angularly averaged over the
only directions perpendicular to the cylinder axis. This result can also
be applied to the out of plane diffuse component of the intensity
collected in grazing incidence small-angle scattering (GISAS)
experiments\cite{sinha88, rauscher95, lazzari02} when the specular region
is avoided so as to make the kinematical approximation satisfactory. See,
in particular, the discussion reported in sect. III.A of \cite{rauscher95}.

Before passing to the illustration of the results relevant to the triangle,
we recall three basic relations worked out in Ref.\cite{cicc09} to which one
should defer for details. They generalize those of the three dimensional(3D)
case and are respectively related to the values at the origin of the first,
second and third derivative of the 2D CF. They read
\begeq\label{elemprop}
\gamma\p(0)= - L/\pi S,
\endeq
\begeq\label{A.12}
\gamma\pp(0^+)=\frac{1}{2\pi S}\sum_{i=1}^M [(\pi-\psi_{i})\cot\psi_{i}\ + 1],
\endeq
and
\begeq\label{B.21}
\gamma\ppp(0^+)\,
 =\,  \frac{1}{4\pi S}\int_L \frac{1}{{R^2}_{curv}(\tau)}d\tau.
\endeq
Here $L$ and $S$ are the perimeter length and the surface area,
the $\psi_{i}$s are the angles at the possible $M$ vertices of the boundary
of the figure and ${R}_{curv}(\tau)$ is the curvature radius at the
point of the boundary with curvilinear coordinate $\tau$.

\section{The triangle case}
According to Ref.\cite{cicc09} the distances that determine the intervals
where $\gamma(r)$ takes an appropriate expression are the lengths of the
sides and the heights of the considered triangle. Hence we must first
establish the order existing among these distances.  To this aim, given
a  triangle, we convey to name $a$ the shortest side, $c$ the longest one
and $b$ the third one so as to always have $a<b<c$. Vertices opposite to
sides $a$, $b$  and $c$  are named  $A$, $B$ and $C$, and the angles at
vertices $A$, $B$ and $C$ as $\alpha$, $\beta$ and $\gamma$  (the context
will avoid any confusion with the CF's symbol $\gamma$). The height from
$A$ to $a$ is $h_a$ and similarly for $h_b$ and $h_c$  (see figure 1).
Owing to condition $a<b<c$ the following inequalities
\begeq\label {b.1}
a\ < \ b \ <\ c,\quad h_c\ <\ h_b\ <\ h_a \quad{\rm and}\quad \alpha\
<\ \beta\ <\ \gamma
\endeq
always hold true.
To order the six lengths: $a,b,c,h_a,h_b$ and $h_c$ we need to know if
$a<h_a$ or $a>h_a$ and if the triangle is acute or obtuse. Hence the four
distinct shapes  shown in Fig. 1.
\begin{figure}[h]
\includegraphics[width=24pc]{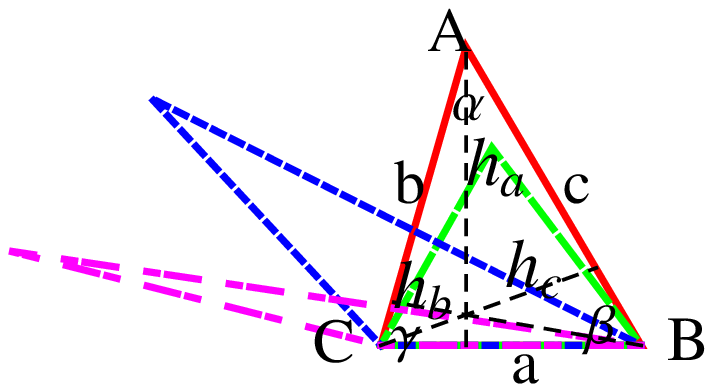}
\caption{\label{fig1} The triangle's four distinct cases $\cA$, $\cB$,
$\cC$ and $\cD$.
Red continuous sides $\to$ shape $\cA$, green dashed sides $\to$ shape
$\cB$, blue short-dashed sides $\to$ shape $\cC$  and magenta dot-dashed
sides $\to$ shape $\cD$. The side lengths of  the shown  $\cA$, $\cB$,
$\cC$ and $\cD$ triangles are (1,\,1.5,\,1.611), (1,\,1.06,\,1.127),
(1,\,1.5,\,2.239) and (1,\,1.5,\,2.470), respectively. }
\end{figure}

\noindent Their names respectively are:
\begin{eqnarray}
\cA\  &\to&\ \ h_c\ <\ h_b\ <\ a\ <\ h_a\ <\ b\ <\ c,\quad
\gamma\,<\,{\pi}/{2}\label{b.2}\\
\cB\  &\to&\ \ h_c\ <\ h_b\ <\ h_a\ <\ a\ <\ b\ <\ c,\quad
\gamma\,<\,{\pi}/{2}\nonumber \\
\cC\  &\to&\ \ h_c\ <\ h_b\ <\ a\ <\ h_a\ <\ b\ <\ c,\quad
\gamma\,>\,{\pi}/{2}\nonumber\\
\cD\  &\to&\ \ h_c\ <\ h_b\ <\ h_a\ <\ a\ <\ b\ <\ c,\quad
\gamma\,>\,{\pi}/{2}.\nonumber
\end{eqnarray}
These inequalities show that we have six $r$-intervals. In case $\cA$
they are: $[0,\,h_c]$, $[h_c,\,h_b]$, $[h_b,\,a]$, $[a,\,h_a]$,
$[h_a,\,b]$ and $[b,\,c]$. They will be referred to as $I\cA,\, II\cA,\,
III\cA,\,IV\cA,\,V\cA$ and $VI\cA$, respectively. The same convention is adopted
in cases $\cB,\,\cC$ and $\cD$. Since  intervals $I\cA$, $I\cB$, $I\cC$ and $I\cD$
coincide we shall simply speak of interval I. The same happens for
intervals II and VI. In each of the subcases listed in Eq.(\ref{b.2})
we can apply the procedure expounded in ref.\cite{cicc09} and determine
the expression of the second derivative of the (isotropic) CF of the
considered triangle. The procedure is straightforward though somewhat
cumbersome. After putting
\begeq\label{d.1}
\Aang \equiv 3+(\pi-\alpha)\cot\alpha + (\pi-\beta)\cot\beta +
(\pi-\gamma)\cot\gamma,
\endeq
\begeq\label{d.2}
\Bang\equiv (\pi-2\alpha)\cot\alpha + (\pi-2\beta)\cot\beta +
2(1+\gamma\cot\gamma),
\endeq
\begeq\label{d.3}
\Cang\equiv 4+2(\pi-2\alpha)\cot\alpha + \pi\cot\beta +
\pi\cot\gamma,
\endeq
\begeq\label{d.3a}
c(\alpha,\beta)\equiv \cot\alpha\,+\,\cot\beta
\endeq
and
\begeq\label{d.4}
\Omega_2(x)\equiv -2x\sqrt{1-x^2}-2\arccos(x),
\endeq
the final expressions of  $\gamma\pp(r)$  are
\begin{eqnarray}
{\gamma\pp}_{_{I}}(r) &=& {\Aang}/({2\pi S}),
\quad\quad\quad\quad\quad
\quad\quad\quad\quad\quad 0<r<h_c;\label{a1}\\  
{\gamma\pp}_{_{II}}(r) &=& [{c(\alpha,\beta)\Omega_2(h_c/r)+\Aang}]
/({2\pi S}), \quad\quad
h_c<r<h_b;            \label{a2}\\     
{\gamma\pp}_{_{III\cA}}(r) &=&
\frac{c(\alpha,\beta)\Omega_2(h_c/r)+
c(\alpha,\gamma)\Omega_2(h_b/r)+\Aang}{2\pi S},\, h_b<r<a;\label{a3}\\    
{\gamma\pp}_{_{IV\cA}}(r) &=& \frac{
c(\alpha,\beta)\Omega_2(h_c/r)+
c(\alpha,\gamma)\Omega_2(h_b/r)+\Cang}{4\pi S},
\,  a<r<h_a;\label{a5}\\  
{\gamma\pp}_{_{V\cA}}(r)&=& \bigl[c(\alpha,\beta)\Omega_2(h_c/r)+
2c(\beta,\gamma)\Omega_2(h_a/r)+ c(\alpha,\gamma)\Omega_2(h_b/r)
\label{a7}\\
\quad  &  & \quad\quad\quad\quad +\Cang\bigr]/({4\pi S})
,\quad\quad\quad
\quad\quad\quad h_a<r<b;\nonumber \\   
{\gamma\pp}_{_{III\cB}}(r) &=& {\gamma\pp}_{_{III\cA}}(r),
\quad\quad\quad\quad\quad
\quad\quad\quad\quad\quad\quad\quad\quad   h_b<r<h_a;\label{aax}\\ 
{\gamma\pp}_{_{IV\cB}}(r) &=& \bigl[c(\alpha,\beta)\Omega_2(h_c/r)+
c(\beta,\gamma)\Omega_2(h_a/r)+c(\alpha,\gamma)\Omega_2(h_b/r)
\label{aa3}\\   \quad & & \quad\quad\quad\quad\quad
+\Aang\bigr]/{2\pi S},
\quad\quad\quad\quad\quad\quad\quad h_a<r<a;\nonumber \\   
{\gamma\pp}_{_{V\cB}}(r) &=& {\gamma\pp}_{_{V\cA}}(r),
\quad\quad\quad\quad
\quad\quad\quad\quad\quad\quad\quad\quad\quad   a<r<b;\label{aa5}\\
{\gamma\pp}_{_{III\cC}}(r) &=&{\gamma\pp}_{_{II}}(r),
\quad\quad\quad\quad
\quad\quad\quad\quad\quad\quad\quad\quad\ \  h_b<r<a;\label{a4}\\
{\gamma\pp}_{_{IV\cC}}(r)&=&{\gamma\pp}_{_{IV\cA}}(r),\quad\quad\quad\quad
\quad\quad\quad\quad\quad
\quad\quad\quad a<r<h_a;\label{a6}\\
{\gamma\pp}_{_{V\cC}}(r)&=&{\gamma\pp}_{_{IV\cC}}(r),\quad\quad\quad\quad
\quad\quad\quad\quad\quad
\quad\quad\quad  h_a<r<b;\label{a8}\\
{\gamma\pp}_{_{III\cD}}(r) &=&{\gamma\pp}_{_{III\cC}}(r),
\quad\quad\quad\quad\quad
\quad\quad\quad\quad\quad\quad\quad\quad  h_b<r<h_a;\label{aa2}\\
{\gamma\pp}_{_{IV\cD}}(r) &=&{\gamma\pp}_{_{II}}(r),
\quad\quad\quad\quad\quad
\quad\quad\quad\quad\quad\quad\quad\quad  h_a<r<a;\label{aa4}\\
{\gamma\pp}_{_{V\cD}}(r) &=&{\gamma\pp}_{_{IV\cA}}(r),
\quad\quad\quad\quad\quad
\quad\quad\quad\quad\quad\quad\quad\quad a<r<b;\label{aa6}\\
{\gamma\pp}_{_{VI}}(r) &=& \frac{
c(\alpha,\gamma)\Omega_2(h_b/r)+c(\beta,\gamma)\Omega_2(h_a/r)+\Bang}
{4\pi S},\   b<r<c.\label{aa7}  
\end{eqnarray}
From these expressions the calculation of $\gamma\p(r)$ is
straightforward after noting that the primitive of $\Omega_2(h/x)$ is
function $\Omega_1(h,x)$
reported below
\begeq\label{d.5}
\Omega_1(h,x)\equiv  \frac{2h\sqrt{x^2-h^2}}{x}-2x\arccos(h/x).
\endeq
In fact, integrating Eq.~(\ref{aa7}) over the interval $[r,\,c]$ with
$b<r<c$ one finds that
\begin{eqnarray}
{\gamma\p}_{_{VI}}(r) &=& \bigl[\Bang(r-c)+
c(\alpha,\gamma)(\Omega_1(h_b,r)-\Omega_1(h_b,c)) +\nonumber\\
\ & & \ c(\beta,\gamma) (\Omega_1(h_a,r)-\Omega_1(h_a,c))\bigr]
/{4\pi S},\quad b < r < c,\nonumber
\end{eqnarray}
where we used the property that $\gamma\p(r)$ is continuous at $r=c$
and therefore equal to zero there. It is now convenient to put
\begeq\label{ome1}
\odme ab \equiv \Omega_1(h_a,r)-\Omega_1(h_a,b).
\endeq
The previous derivative expression  then reads
\begin{eqnarray}
{\gamma\p}_{_{VI}}(r) &=& \bigl[\Bang(r-c)+
c(\alpha,\gamma)\odme bc  + c(\beta,\gamma)\times \nonumber \\
\ & & \quad \odme ac \bigr]/({4\pi S}),\quad\quad   b < r < c,\label{dd.1}
\end{eqnarray}
\noindent In the same way, integrating Eq.~(\ref{a7})
over the interval $[r,\,b]$ with  $h_a<r<b$ and imposing  the continuity
at $r=b$, one finds that
\begin{eqnarray}
{\gamma\p}_{_{V\cA}}(r)& = &\frac{1}{4\pi S} \bigl[2c(\beta,\gamma)\odme ab
+c(\alpha,\gamma)\odme bb +  \label{dd.2}\\
\  &  & \quad c(\alpha,\beta)\odme cb +  \Cang(r-b)
 \bigr]+{\gamma\p}_{_{VI}}(b),\quad h_a < r < b,\nonumber
\end{eqnarray}

\begin{figure}[!h]
\begin{minipage}{12pc}
\includegraphics[width=12pc]{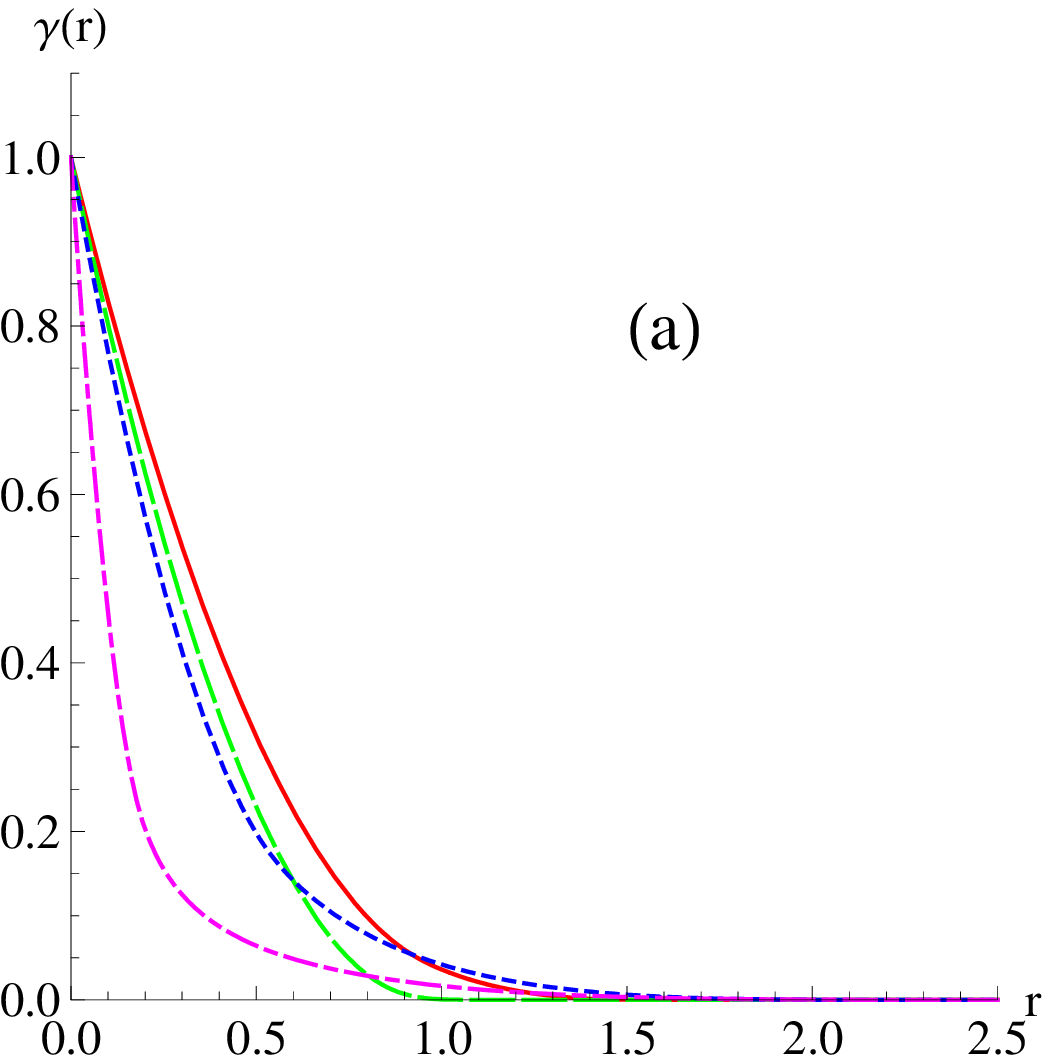}
\end{minipage}
\hspace{2pc}%
\begin{minipage}{12pc}
\includegraphics[width=12pc]{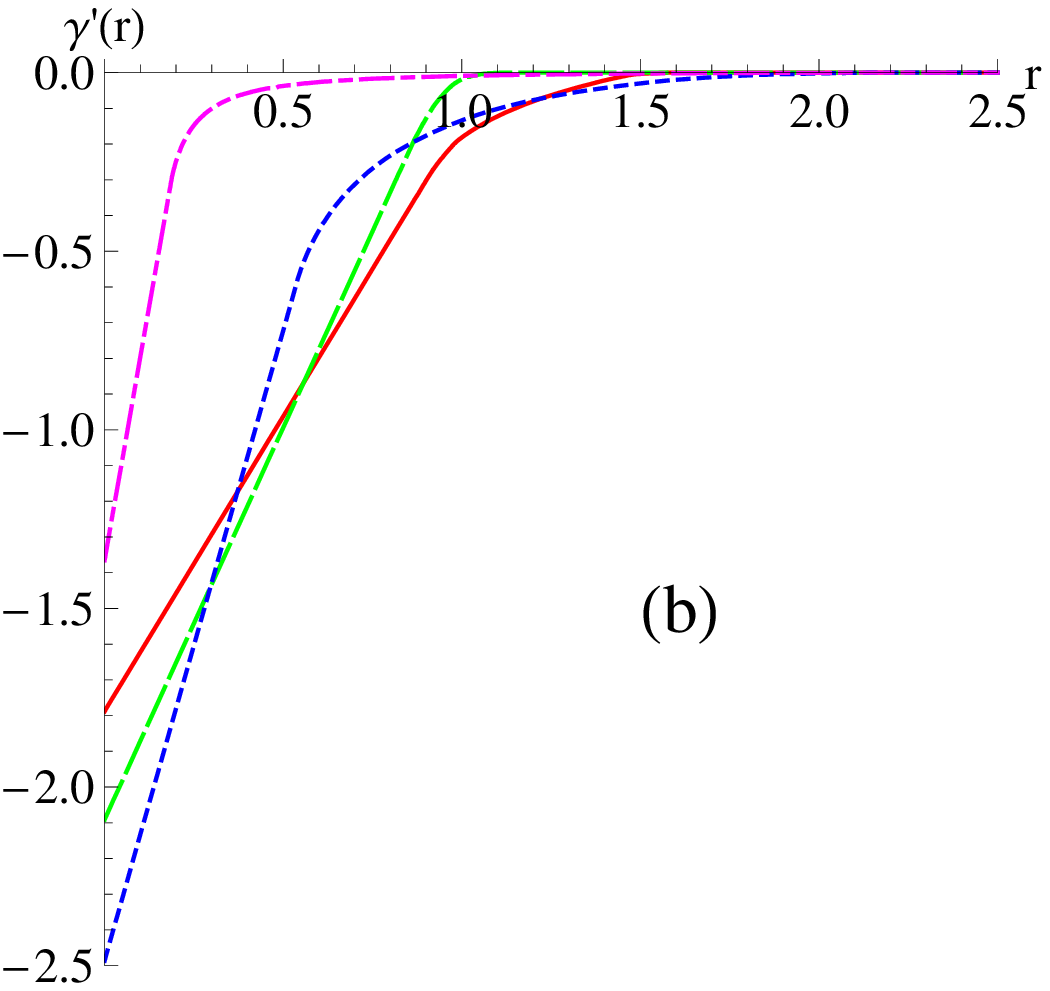}
\end{minipage}
{
\begin{minipage}{12pc}
\includegraphics[width=12pc]{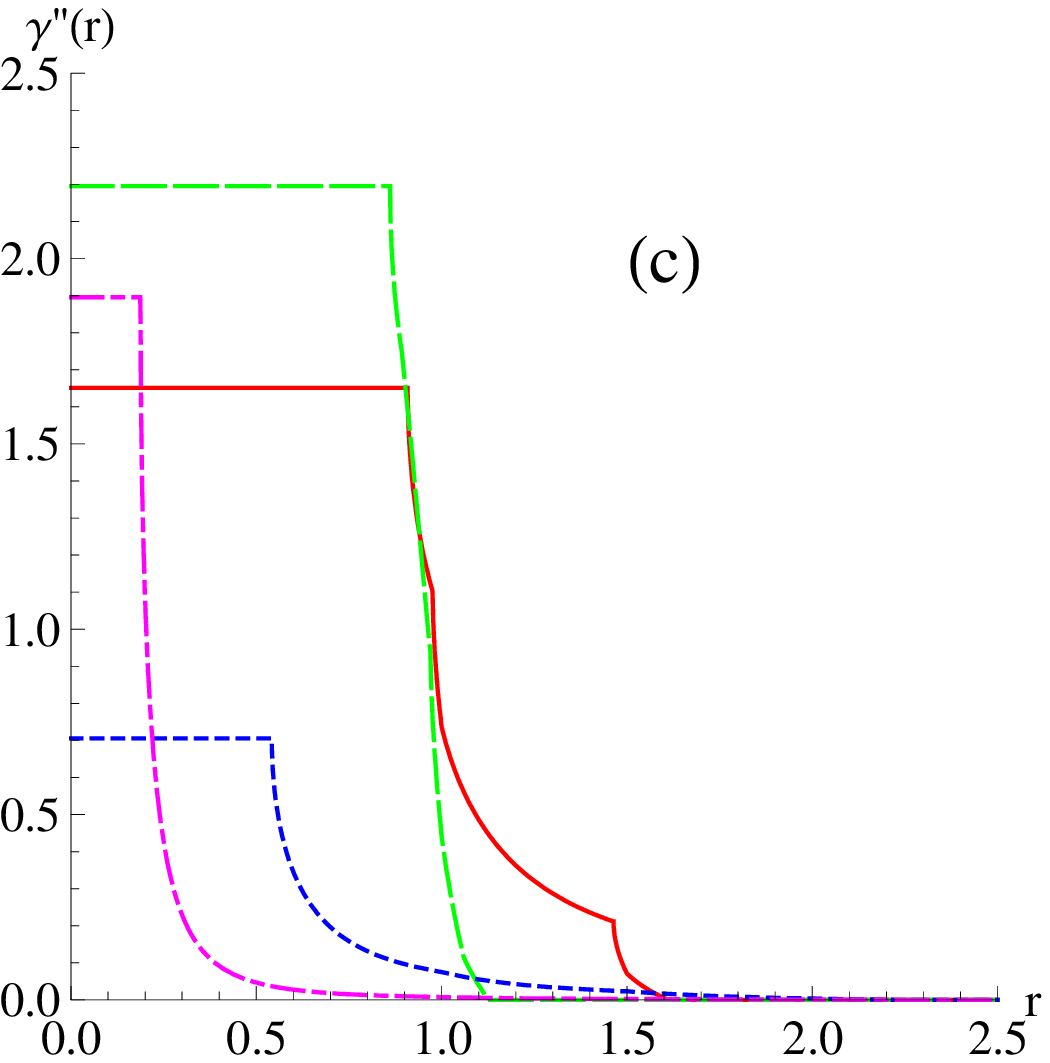}
\end{minipage}
\hspace{2pc}%
\begin{minipage}{12pc}
\includegraphics[width=12pc]{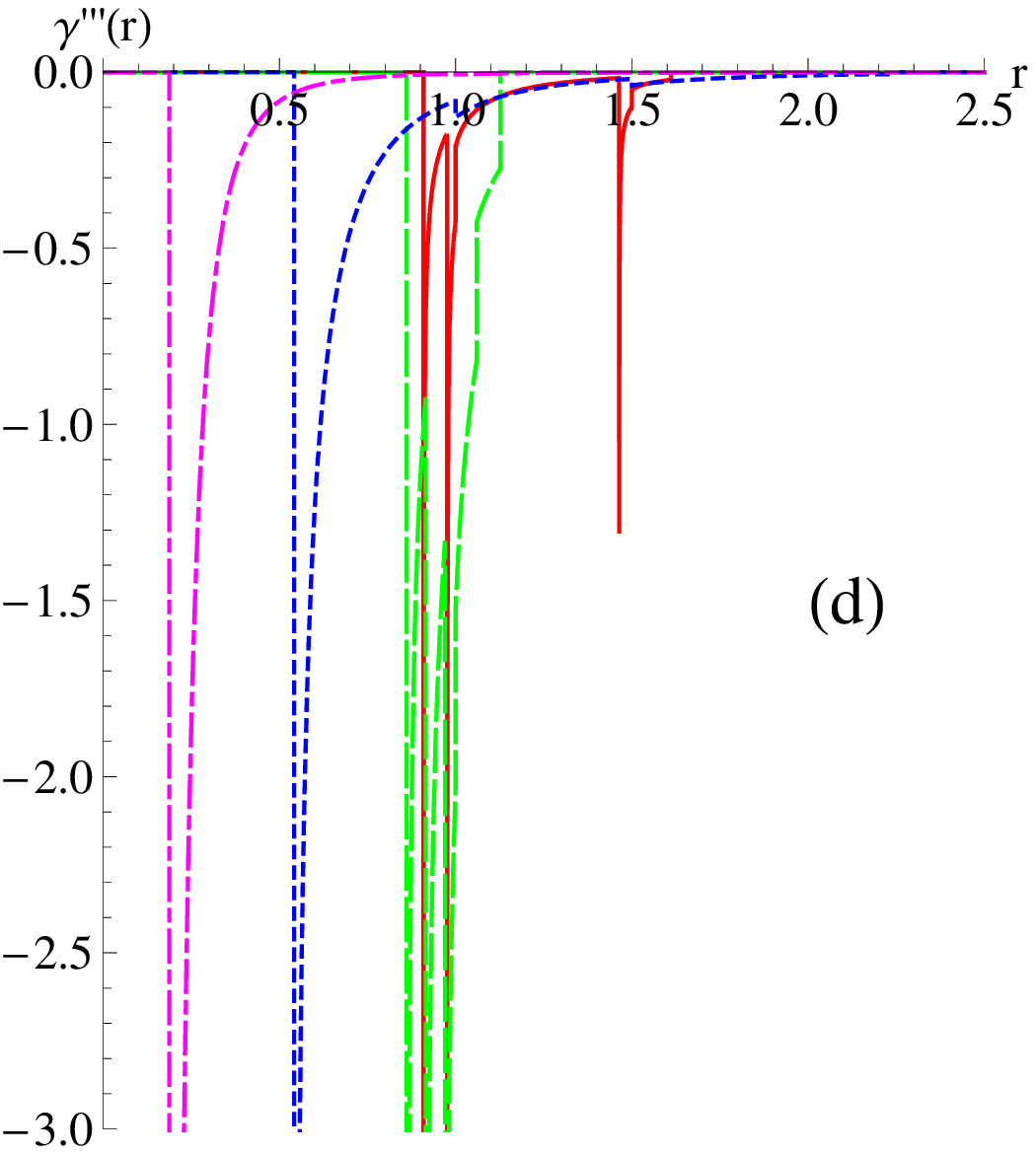}
\end{minipage}
}
\caption {\label{Fig2} {\bf (a)}: plots of $\gamma(r)$ for triangle
$\cA$ (continuous red curve), $\cB$ (green dashed curve), $\cC$
(blue short-dashed) and $\cD$ (magenta dot-dashed). The convention
in the curve drawing is the same  of Fig. 1;  {\bf (b)}: plots of
$\gamma\p(r)$ (note, however, that $\gamma\p(r)$ is  scaled by a factor
$1/5$ in case $\cD$);  {\bf (c)}: plots of $\gamma\pp(r)$ (the shown
$\gamma\pp(r)$s are scaled by factors $1/5$ and $1/15$ in cases $\cC$
and $\cD$, respectively); {\bf (d)}: plots of $\gamma\ppp(r)$ (the
shown $\gamma\ppp(r)$s are scaled by factors $1/20$, $1/5$, $1/10$ and
$1/20$ in cases A, B, C and D).}
\end{figure}

\noindent and then, step by step, that
\begin{eqnarray}
{\gamma\p}_{_{IV\cA}}(r)& = &\frac{1}{4\pi S}\bigl[c(\alpha,\gamma)
\odme b{h_a}+c(\alpha,\beta)\odme c{h_a} +\label{dd.3}\\
\  &  & \quad \Cang(r-h_a)\bigr]  +{\gamma\p}_{_{V\cA}}(h_a),\quad a < r < h_a;\nonumber\\
{\gamma\p}_{_{III\cA}}(r)& = &\frac{1}{2\pi S}\bigl[c(\alpha,\gamma)\odme ba
+ c(\alpha,\beta)\odme ca  +\nonumber \\
\  & &\quad \Aang(r-a)\bigr]
 +  {\gamma\p}_{_{IV\cA}}(a),
\quad\quad\quad h_b < r < a;\label{dd.4}\\
{\gamma\p}_{_{II\cA}}(r)& = &\frac {c(\alpha,\beta)\odme cb +\Aang(r-h_b)}
{2\pi S} +  \label{dd.5}\\
\ & & \quad   {\gamma\p}_{_{III\cA}}(h_b),\quad\quad\quad
\quad\quad\quad h_c < r < h_b;\nonumber\\
{\gamma\p}_{_{I\cA}}(r)& = &\frac{\Aang(r-h_c)}{2\pi S}+
{\gamma\p}_{_{II\cA}}(h_c),\quad\quad 0 < r < h_c.\label{dd.6}
\end{eqnarray}
The expressions of $\gamma\p(r)$ are similarly obtained  in the other
three cases and are reported in the appendix. In the same way, from the
resulting expressions of the first derivative, using the fact that
a primitive of $(\Omega_1(h,x)$ is
\begeq\label{d.6}
\Omega(h,x)\equiv  -{\pi x^2}/{2}+3h\sqrt{x^2-h^2}+(x^2+2h^2)\arcsin(h/x),
\endeq
starting from the outermost $r$-interval and imposing the continuity at the
outermost end points one obtains the algebraic expression of the CF of a
generic triangle. For instance, the CF expression in the outermost interval
is
\begin{eqnarray}\nonumber
{\gamma}_{_{VI}}(r)& =& \Bigl[\Bang(r-c)^2/2+
c(\alpha,\gamma)[\Omega(h_b,r)-\Omega(h_b,c) -\Omega_1(h_b,c)\times 
\nonumber\\
\ &   & (r-c) ] +  c(\beta,\gamma)
[\Omega(h_a,r)-\Omega(h_a,c)-\Omega_1(h_a,c)(r-c))]\Bigr]/{4\pi S}.
\ \label{d.6}\quad\quad 
\end{eqnarray}
This expression, after putting
\begeq\label{A}
\ome ab \equiv \Omega(h_a,r)-\Omega(h_a,b)-\Omega_1(h_a,b)(r-b),
\endeq
simplifies into
\begin{eqnarray}
{\gamma}_{_{VI}}(r)& =& [\Bang(r-c)^2/2+
c(\alpha,\gamma)\ome bc +\label{ee.1}\\
\ & & c(\beta,\gamma) \ome ac ]/({4\pi S}),
\quad\quad\quad  b<r<c.\nonumber
\end{eqnarray}
In the remaining interval,   for case $\cA$ one finds
\begin{eqnarray}
{\gamma}_{_{V\cA}}(r)& = & \frac{1}{4\pi S}\Bigl[\frac{\Cang(r-b)^2}{2}+
c(\alpha,\beta)\ome cb + c(\alpha,\gamma)\ome bb  + \quad\quad  \label{ee.2}\\
\quad & &  \ 2 c(\beta,\gamma) \ome ab \Bigr]+
{\gamma\p}_{_{VI}}(b)(r-b)+{\gamma}_{_{VI}}(b),
\quad\quad h_a<r<b,\nonumber\\
{\gamma}_{_{IV\cA}}(r)& = & \frac{1}{4\pi S}\Bigl[\frac{\Cang(r-h_a)^2}{2}+
c(\alpha,\gamma)\ome b{h_a}+c(\alpha,\beta)\times \label{ee.3}\\
\quad  &   & \ome c{h_a}\Bigr] +{\gamma\p}_{_{V\cA}}(h_a)(r-h_a) +{\gamma}_{_{V\cA}}(h_a),
\quad\quad\quad a<r<h_a,\nonumber\\
{\gamma}_{_{III\cA}}(r)& =  & \frac{1}{4\pi S} \Bigl[\frac{\Aang(r-a)^2}{2}+
c(\alpha,\gamma)\ome ba   + c(\alpha,\beta)\ome ca  \Bigr]\label{ee.3}\\
\quad  &   &\quad + {\gamma\p}_{_{IV\cA}}(a)(r-a) +{\gamma}_{_{IV\cA}}(a),
\quad\quad\quad\quad\quad\quad\quad\quad h_b<r<a,\nonumber\\
{\gamma}_{_{II\cA}}(r)& = &\frac{1}{2\pi S}\Bigl[{\frac{\Cang(r-h_b)^2}{2}+
 c(\alpha,\beta)\ome c{h_b} }\Bigr] + \label{ee.5}\\
\quad  &   & \quad\quad
 {\gamma\p}_{_{III\cA}}(h_b)(r-h_b)+{\gamma}_{_{III\cA}}(h_b),
\quad\quad\quad\quad\quad\quad\quad h_c<r<h_b,\nonumber\\
{\gamma}_{_{I\cA}}(r)& = & \frac{\Aang(r-h_c)^2}{4\pi S} +
{\gamma\p}_{_{II\cA}}(h_c)(r-h_c)+{\gamma}_{_{II\cA}}(h_c),\ 0<r<h_c.\label{ee.6}
\end{eqnarray}
The expressions relevant to the other cases are given in the appendix.
Figures 2a-d show the plots of $\gamma(r)$ and its first three
derivatives for the four triangle cases. The CFs look each other rather
similar because the different supports cannot be appreciated on the
figure scale. The first derivatives show greater difference since their
behaviours about $r=0$ signal that the specific length (i.e. the ratio
perimeter/surface, see Eq.(\ref{elemprop})) increases as one passes
from case $\cA$ to $\cD$ (note that the magenta curve is scaled). The
differences become more evident for the second derivatives and
dramatic for the third one. Here the cuspids signal algebraic
singularities $\propto (r-h)^{-1/2}$ and they involve the only heights
which meet the opposite  side within the triangle. This explains why
Fig. 3d shows three spikes in cases $\cA$ and $\cB$ and only one spike
in cases $\cC$ and $\cD$.

\section{Conclusions}
We have briefly described the cases where the algebraic knowledge of the
CF of a plane figure can usefully be applied to analyze observed scattering
intensities. In particular we have reported here the explicit expression
of the CF of a general triangle. Besides their intrinsic interest for the
shown singularities, which are important from the point of view of the
figure reconstruction,  these expressions allow us to evaluate the CF of
the associated cylinder by a simple quadrature.

\appendix{\section{Expressions for cases $\cB$, $\cC$ and $\cD$}}
For completeness we give the expressions of the CF and its 1st
derivative in remaining cases $\cB$, $\cC$ and $\cD$. For the first
derivative one finds
\begin{eqnarray}
\gamma\p_{_{VI\cB}}(r) &=&\gamma\p_{_{VI}}(r),
\quad\quad\quad\quad\quad\quad\quad\quad\quad\quad\quad
\quad\quad b<r<c;\quad\quad\quad\quad\label{bd6}\quad\quad\quad\quad\quad\quad\\
\gamma\p_{_{V\cB}}(r) &=&\frac{1}{4 \pi S}\Bigl[2 {c} (\beta,\gamma)
\odme ab  + {c} (\alpha , \gamma) \odme bb + {c} (\alpha , \beta )\times
\label{bd5}\\
\quad &  &  \odme cb  + (r - b)
{\cC} (\alpha , \beta , \gamma )\Bigr] + \gamma\p_{_{VI}}(b),
\quad\quad\quad\quad\quad  a<r<b;\nonumber\\
\gamma\p_{_{IV\cB}}(r) &=&   \frac{1}{2 \pi S}\Bigl[
c(\alpha, \beta)\odme ca +  c(\beta, \gamma)\odme aa + c(\alpha, \gamma)
\label{bd4}\\
\quad & &  \quad \odme ba
+\cA(\alpha, \beta, \gamma)(r - a) \Bigr] + \gamma\p_{_{V\cB}}(a),
\quad\quad h_a<r<a;\nonumber\\
\gamma\p_{_{III\cB}}(r) &=& \frac{1}{2 \pi S}\Bigl[
 c(\alpha, \beta) \odme c{h_a} +c(\alpha, \gamma) \odme b{h_a} +
 \label{bd3}\\ \nonumber\\
 \ & &   \quad {\cA}(\alpha,\beta,\gamma )(r - h_a) \Bigr] + \gamma\p_{_{IV\cB}}(h_a),\
\quad\quad\quad\quad\quad\quad\quad h_b<r<h_a;
 \nonumber\\
\gamma\p_{_{II\cB}}(r) &=& \frac{c(\alpha, \beta)\left(\Omega_1(h_c,r)-
\Omega_1(h_c,h_b)\right) +\cA(\alpha, \beta, \gamma)(r -h_b)}{2 \pi S}
 \label{bd2}  \\
\ & &  \quad \quad + \gamma\p_{_{II\cB}}(h_b),
\quad\quad\quad\quad\quad\quad\quad\quad\quad\quad\quad\quad h_c<r<h_b;\nonumber   \\
\gamma\p_{_{I\cB}}(r) &=& \frac{\cA(\alpha, \beta, \gamma)
(r -h_c)}{2 \pi S} + \gamma\p_{_{II\cB}}(h_c),
 \quad\quad\quad\quad\quad
 \quad 0 < r < h_c;\label{bd1}
\end{eqnarray}
in case $\cB$,
\begin{eqnarray}
\gamma\p_{_{VI\cC}}(r) &=&\gamma\p_{_{VI}}(r),\quad\quad\quad
\quad\quad\quad\quad\quad\quad\quad\quad\quad\quad\quad\quad
\quad\quad b<r<c;\quad\quad\quad\quad\quad\quad\quad\quad\label{cd6}\\
\gamma\p_{_{V\cC}}(r)&=&\frac{1}{4 \pi S}\bigl[
c(\alpha, \gamma)\odme bb  + c(\alpha, \beta)\odme cb
+\cC(\alpha, \beta, \gamma)(r - b)\bigr] \label{cd5}\\
\ & &\quad\quad +  \gamma\p_{_{VI\cC}}(b),\quad\quad\quad
\quad\quad\quad\quad\quad\quad\quad\quad\quad h_a<r<b;\nonumber\\
\gamma\p_{_{IV\cC}}(r)&=&\frac{1}{4 \pi S}\bigl[
c(\alpha, \gamma)\odme b{h_a}  +
c(\alpha, \beta)\odme c{h_a} +\label{cd4}\\
\ & &\quad\cC(\alpha, \beta, \gamma)(r -h_a)\bigr]
+\gamma\p_{_{V\cC}}(h_a),\quad\quad\quad
\quad\quad\quad a<r<h_a;\nonumber\\
\gamma\p_{_{III\cC}}(r)&=&\frac{c(\alpha, \beta)\odme ca +
\cA(\alpha, \beta, \gamma)(r - a)}{2 \pi S}  + \gamma\p_{_{IV\cC}}(a),
\quad h_b <r<a; \label{cd3}\\
\gamma\p_{_{II\cC}}(r) &=&
\frac{c(\alpha, \beta)\odme c{h_b}  + \cA(\alpha, \beta,
\gamma)(r -h_b)}{2 \pi S}+\gamma\p_{_{II\cC}}(h_b),
\ \ h_c<r<h_b;\label{cd2}\\
\gamma\p_{_{I\cC}}(r) &=& \frac{\cA(\alpha, \beta, \gamma)
(r -h_c)}{2 \pi S}
 + \gamma\p_{_{II\cC}}(h_c),\quad\quad\quad\quad\quad
 \quad\quad\quad\quad 0 < r < h_c;\label{cd1}
\end{eqnarray}
in case $\cC$, and
\begin{eqnarray}
\gamma\p_{_{VI\cD}}(r) &=&\gamma\p_{_{VI}}(r),\quad\quad\quad
\quad\quad\quad\quad\quad\quad\quad\quad\quad\quad\quad\quad
\quad\quad b<r<c;\quad\quad\quad\quad\label{dd6}\\
\gamma\p_{_{V\cD}}(r)&=&
\frac{1}{4 \pi S}\bigl[ c(\alpha, \gamma)\odme bb +
c(\alpha, \beta)\odme cb + \label{dd5}\\
\quad  & &  \quad \cC(\alpha, \beta, \gamma)(r - b)\bigr]+\gamma\p_{_{VI\cD}}(a),
\quad\quad\quad\quad \quad\quad
\quad\quad\quad  a<r<b;\quad\quad\nonumber\\
\gamma\p_{_{IV\cD}}(r)&=&
\frac{\cA(\alpha,\beta,\gamma)(r-a)+c(\alpha,\beta)\odme ca }
{2 \pi S}
+\gamma\p_{_{V\cD}}(a),\quad h_a<r<a; \quad\quad\label{dd4}\\
\gamma\p_{_{III\cD}}(r) &=&
\frac{\cA\left(\alpha, \beta, \gamma\right)(r-h_a)+c(\alpha,\beta)
+\odme c{h_b} }{2 \pi S}+\label{dd3}\\
\ & & \quad \gamma\p_{_{IV\cD}}(h_a),\quad\quad\quad\quad\quad\quad
\quad\quad\quad\quad\quad\quad\quad\quad\quad  h_b<r<h_a;\quad\nonumber \\
\gamma\p_{_{II\cD}}(r) &=&
\frac{c(\alpha, \beta)\odme c{h_b}  + \cA(\alpha, \beta, \gamma)
(r -h_b)}{2 \pi S}
+  \gamma\p_{_{II\cD}}(h_b),\ h_c<r<h_b;\quad\quad \label{dd2}\\
\gamma\p_{_{I\cD}}(r) &=& \frac{\cA(\alpha, \beta, \gamma)(r -h_c)}
{2 \pi S}
 + \gamma\p_{_{II\cD}}(h_c),\quad\quad\quad
 \quad\quad\quad\quad\quad\ 0 < r < h_c\label{dd1}
\end{eqnarray}
in case $\cD$.

\noindent For the CF one finds
\begin{eqnarray}
\gamma_{_{VI\cB}}(r) &=&\gamma_{_{VI}}(r),\quad\quad\quad b<r<c\label{b6}\\
\gamma_{_{V\cB}}(r) &=&\frac{1}{4 \pi S}\Bigl[
2{c}(\beta,\gamma)\ome ab  + {c}(\alpha, \gamma) \ome bb   +
{c}(\alpha,\beta)\ome cb    \label{b5}\\
\quad & &\quad\quad
+{\cC}(\alpha , \beta , \gamma )\frac{(r - b)^2}{2}\Bigr]
  + \gamma\p_{_{VI}}(b)(r-b) + \gamma_{_{VI}}(b),
\quad \quad  a<r<b;\nonumber \\
\gamma_{_{IV\cB}}(r) &=&\frac{1}{2 \pi S}\Bigl[
{c}(\beta,\gamma)\ome aa  +
{c}(\alpha, \gamma) \ome ba   +
{c}(\alpha,\beta)\ome ca   \label{b4}\\
\quad & &\quad
+{\cA}(\alpha , \beta , \gamma )\frac{(r - a)^2}{2}\Bigr] +
\gamma\p_{_{V\cB}}(a)(r-a) +
\gamma_{_{V\cB}}(a),
\quad \quad  h_a<r<a;\nonumber \\
\gamma_{_{III\cB}}(r) &=&\frac{1}{2 \pi S}\Bigl[
{c}(\alpha,\beta)\ome c{h_a} +{c}(\alpha, \gamma) \ome b{h_a}
+{\cA}(\alpha,\beta,\gamma )\frac{(r - h_a)^2}{2}\Bigr]
\label{b3}\\
\quad & &\quad\quad + \gamma\p_{_{IV\cB}}(h_a)(r-h_a) +
\gamma_{_{IV\cB}}(h_a),\quad\quad\quad
\quad\quad\quad \quad  h_b<r<h_a;\nonumber \\
\gamma_{_{II\cB}}(r) &=&\frac{1}{2 \pi S}\Bigl[{c}(\alpha,\beta)
\ome c{h_b} + {\cA}(\alpha , \beta , \gamma )\frac{(r - h_b)^2}{2}\Bigr]
\label{b2}\\
\quad & &\quad\quad + \gamma\p_{_{III\cB}}(h_b)(r-h_b) +
\gamma_{_{III\cB}}(h_b),\quad\quad
\quad\quad\quad\quad\quad h_c<r<h_b;\quad\quad\quad\quad 
\nonumber \\
\gamma_{_{I\cB}}(r) &=&\frac{{\cA}(\alpha , \beta , \gamma )
{(r - h_c)^2}}{4 \pi S}
 +\gamma\p_{_{II\cB}}(h_c)(r-h_c) +\gamma_{_{II\cB}}(h_c),
\quad   0<r<h_c \quad\label{b2}
\end{eqnarray}
in case $\cB$,
\begin{eqnarray}
\gamma_{_{VI\cC}}(r) &=&\gamma_{_{VI}}(r),\quad\quad\quad\quad
\quad\quad \quad\quad\quad \quad\quad\quad\quad\quad\quad\quad
\quad\quad\quad\quad\quad b<r<c;\quad\quad\quad\quad\label{c6}\\
\gamma_{_{V\cC}}(r) &=&\frac{1}{4 \pi S}\Bigl[
{c}(\alpha,\beta)\ome cb + {c}(\alpha, \gamma) \ome bb
+{\cC}(\alpha , \beta , \gamma ){(r - b)^2}/{2}\Bigr]\label{c5}\\
\quad & &\quad\quad + \gamma\p_{_{VI}}(b)(r-b) + \gamma_{_{VI}}(b),
\quad\quad\quad\quad\quad\quad\quad\quad  h_a<r<b;\nonumber \\
\gamma_{_{IV\cC}}(r) &=&\frac{1}{4 \pi S}\Bigl[
{c}(\alpha,\beta)\ome c{h_a} +{c}(\alpha, \gamma) \ome b{h_a}
+{\cC}(\alpha , \beta , \gamma ){(r - h_a)^2}/{2}\Bigr]\label{c4}\\
\quad & &\quad\quad
+ \gamma\p_{_{V\cC}}(h_a)(r-h_a) +\gamma_{_{V\cC}}(h_a),
\quad\quad\quad\quad\quad\quad\quad \quad  a<r<h_a;\nonumber \\
\gamma_{_{III\cC}}(r) &=&\frac{1}{2 \pi S}\Bigl[
{c}(\alpha,\beta)\ome ca  +{\cA}(\alpha,\beta,\gamma)
\frac{(r - a)^2}{2}\Bigr]\label{c3}\\
\quad & &\quad\quad + \gamma\p_{_{IV\cC}}(a)(r-a) +\gamma_{_{IV\cC}}(a),
\quad\quad\quad\quad\quad\quad\quad\quad  h_b<r<a;\nonumber\\
\gamma_{_{II\cC}}(r) &=&\frac{1}{2 \pi S}\Bigl[
{c}(\alpha,\beta)\ome c{h_b} +
{\cA}(\alpha , \beta , \gamma )\frac{(r - h_b)^2}{2}\Bigr]\label{c2}\\
\quad & &\quad\quad +\gamma\p_{_{III\cC}}(h_b)(r-h_b) +
\gamma_{_{III\cC}}(h_b),\quad
\quad \quad\quad\quad\quad  h_c<r<h_b;\nonumber \\
\gamma_{_{I\cC}}(r) &=&\frac{{\cA}(\alpha , \beta , \gamma )
{(r - h_c)^2}}{4 \pi S}
 +\gamma\p_{_{II\cC}}(h_c)(r-h_c) +\gamma_{_{II\cC}}(h_c),
\quad \quad  0<r<h_c\label{c1}
\end{eqnarray}
in case $\cC$, and
\begin{eqnarray}
\gamma_{_{VI\cD}}(r) &=&\gamma_{_{VI}}(r),\quad\quad\quad\quad
\quad\quad\quad\quad\quad\quad\quad\quad\quad\quad\quad\quad
\quad\quad\quad\quad b<r<c;\label{d6}\quad\quad\quad\quad\quad\quad\\
\gamma_{_{V\cD}}(r) &=&\frac{1}{4 \pi S}\Bigl[{c}(\alpha,\beta)
\ome cb  +
{c}(\alpha,\gamma)\ome bb  +{\cC}(\alpha,\beta,\gamma)
\frac{(r - b)^2}{2}\Bigr]\label{d5}\\
\quad & &\quad\quad + \gamma\p_{_{VI}}(b)(r-b)+
\gamma_{_{VI}}(b),\quad\quad\quad\quad
\quad\quad\quad\quad\quad \quad  a<r<b;\nonumber \\
\gamma_{_{IV\cD}}(r) &=&\frac{1}{2 \pi S}\Bigl[{c}
(\alpha,\beta)\ome ca +
{\cA}(\alpha,\beta,\gamma)\frac{(r - a)^2}{2}\Bigr]\label{d4}\\
\quad & &\quad\quad +\gamma\p_{_{V\cD}}(a)(r-a) +\gamma_{_{V\cD}}(a),
\quad\quad\quad\quad
\quad\quad \quad\quad\quad\quad  h_a<r<a;\nonumber \\
\gamma_{_{III\cD}}(r) &=&\frac{1}{2 \pi S}\Bigl[{c}(\alpha,\beta)
\ome c{h_a}  +{\cA}(\alpha,\beta,\gamma )\frac{(r - h_a)^2}{2}
\Bigr]\label{d3}\\
\quad & &\quad\quad +\gamma\p_{_{IV\cD}}(h_a)(r-h_a) +
\gamma_{_{IV\cD}}(h_a),\quad\quad\quad
\quad\quad  \quad\quad\quad h_b<r<h_a;\nonumber \\
\gamma_{_{II\cD}}(r) &=&\frac{1}{2 \pi S}\Bigl[{c}(\alpha,\beta)
\ome c{h_b} +
{\cA}(\alpha,\beta,\gamma)\frac{(r - h_b)^2}{2}\Bigr]\label{d2}\\
\quad & &\quad\quad +\gamma\p_{_{III\cD}}(h_b)(r-h_b) +
\gamma_{_{III\cD}}(h_b), \quad\quad
\quad\quad\quad\quad\quad\quad  h_c<r<h_b;\nonumber \\
\gamma_{_{I\cD}}(r) &=&\frac{{\cA}(\alpha , \beta , \gamma )
{(r - h_c)^2}}{4 \pi S}
 +\gamma\p_{_{II\cD}}(h_c)(r-h_c) +\gamma_{_{II\cD}}(h_c),
\quad \quad  0<r<h_c\label{d1}
\end{eqnarray}
in case $\cD$.
\\  
\\  
\\
  \noindent
{\bfdodc Acknowledgments}
\\ 
\\ 
\noindent
I thank Dr Wilfried Gille for his critical reading of the manuscript 
and for having checked most of the above formulae.
\vfill\eject

\end{document}